\documentclass[fleqn,twoside]{article}
\usepackage{espcrc2,amsmath2000,amsfonts,amssymb,epic,eepic}

\usepackage[figuresright]{rotating}

\newcommand{\1}{{\sf 1 \!\! 1}}
\newcommand{\bra}[1]{\left\langle#1\right|}
\newcommand{\ket}[1]{\left|#1\right\rangle}
\newcommand{\mat}[4]{\left(%
    \begin{array}{cc}#1&#2 \\#3&#4\end{array}\right)}

\newcommand{\blm}[1]{$#1$}

\newcommand{\stanlink}[1]{{\begin{picture}(50,15)(-5,-2)%
      \put(18,-2.17){\small$#1$}%
      \path(0,0)(40,0)%
      \put(0,0){$\circle*{3}$}%
      \put(40,0){$\circle*{3}$}%
      \put(-3,-11){\small $x$}%
      \put(27,-11){\small $x\!+\!\hat\mu$}%
    \end{picture}}}

\newcommand{\plaq}[1]{{\begin{picture}(45,30)(-7,10)%
      \ifthenelse{\equal{#1}{b}}{\path(5,0)(25,0)}{}%
      \ifthenelse{\equal{#1}{r}}{\path(30,5)(30,25)}{}%
      \ifthenelse{\equal{#1}{t}}{\path(5,30)(25,30)}{}%
      \ifthenelse{\equal{#1}{l}}{\path(0,5)(0,25)}{}%
      \ifthenelse{\equal{#1}{a}}{\path(5,0)(25,0)\path(30,5)(30,25)%
        \path(5,30)(25,30)\path(0,5)(0,25)}{}%
      \put(0,0){$\circle*{5}$}%
      \put(30,0){$\circle*{5}$}%
      \put(0,30){$\circle*{5}$}%
      \put(30,30){$\circle*{5}$}%
    \end{picture}}}

\makeatletter

\newbox\slashbox \setbox\slashbox=\hbox{\large$/$}
\def\pslash#1{\setbox\@tempboxa=\hbox{$#1$}
  \@tempdima=0.5\wd\slashbox \advance\@tempdima 0.5\wd\@tempboxa
  \copy\slashbox \kern-\@tempdima \box\@tempboxa}

\def\openone{\leavevmode\hbox{\small\texttt{1}\kern-6pt\large\texttt{1}}}

\newcommand{\lsim}{
 \mathrel{\setbox0=\hbox{$<$}\raise0.6ex\copy0\kern-\wd0
 \lower0.65ex\hbox{$\sim$}}}

\DeclareMathOperator{\re}{Re}
\DeclareMathOperator{\tr}{tr}
\DeclareMathOperator{\diag}{diag}
\DeclareMathOperator{\Sdet}{Sdet}
\DeclareMathSymbol{\lozenge} {\mathord}{AMSa}{"07}
\DeclareMathSymbol{\lesssim}      {\mathrel}{AMSa}{"2E}

{\renewcommand{\labelitemi}{#1}
  \begin{itemize}\itemsep -1mm}
{\end{itemize}}

{\renewcommand{\labelenumi}{#1\theenumi.}
  \begin{enumerate}\itemsep -1mm}
{\end{enumerate}}

\def\thickhline{%
  \noalign{\ifnum0=`}\fi\hrule \@height 3pt \futurelet
   \reserved@a\@xhline}

\title{The Color-Flavor Transformation and Lattice QCD\thanks{ 
       Based on two talks by the authors at Lattice 2002.}}

\author{B. Schlittgen\address[BS]{Department of Physics, Yale University,
        New Haven, CT 06520-8120, USA}
        and
        T. Wettig\addressmark[BS]$^{,}$\address{RIKEN-BNL Research Center, 
                Upton, NY, 11973-5000, USA}}

\begin{document}

\begin{abstract}
We present the color-flavor transformation for gauge group SU($N_c$)
and discuss its application to lattice QCD.
\vspace{-2mm}
\end{abstract}

\maketitle

\section{Introduction}

Motivated by the study of disordered systems in condensed
matter physics, 
the color-flavor transformation was first derived by Zirnbauer
in 1996 for an integral over U($N_c$) \cite{Zirn96},
\begin{multline}
\int\limits_{\text{U}(N_c)}\!\!\!dU \exp\left(
    \bar{\psi}_{{ x+\hat\mu,}{ a}}^{ i}\,{ U^{ij}}
    \psi_{{ x,}{ a}}^{ j}+ 
    \bar{\psi}_{{ x,}{ b}}^{ j}\,{U^{\dagger ji}}
    \psi_{{ x+\hat\mu,}{ b}}^{ i}\right) \\
  =\int D\mu_{N_c}({ Z},{ \tilde{Z}})\\
   \times\exp\left( 
    \bar{\psi}_{{ x+\hat\mu,}{ a}}^{ i}{ Z_{ab}}
    \psi_{{ x+\hat\mu,}{ b}}^{ i}+
    \bar{\psi}_{{ x,}{ b}}^{ i}
    { \tilde{Z}_{ba}}\psi_{{ x,}{ a}}^{ i} \right).
\label{eq:uncolfla}
\end{multline}
In the above equation, $\psi$ and $\bar\psi$ are $\mathbb{Z}_2$-graded
tensors, $i,j=1,\ldots,N_c$ are color indices, $a=1,\ldots,n_+$ and
$b=1,\ldots,n_-$ are flavor indices, and the supermatrices $Z$ and
$\tilde Z$ parameterize the coset space
U$(n_{+}+n_{-}|n_{+}+n_{-})/$U$(n_{+}|n_{+})\times$U$(n_{-}|n_{-})$.
Their boson-boson parts satisfy $\tilde Z_{\rm BB}= Z_{\rm BB}^\dagger$,
and for the fermion-fermion parts, $\tilde Z_{\rm FF}=- Z_{\rm
  FF}^\dagger$.  The measure is given by $D\mu_{N_c}(Z,\tilde{Z})=
d(Z,\tilde{Z})$ $\Sdet(\1-\tilde ZZ)^{N_c}$.  For further progress,
see \cite{cftprogress}.

Notice that before the transformation, the color indices of $\psi$ and
$\bar\psi$ are coupled by the matrix $U$, while the flavor indices are
diagonal. In the transformed integral, the flavor indices are coupled
by the matrix $Z$, and the color indices are diagonal --- thus the
name of the transformation.

The integral over the gauge field closely resembles the one for a
single link in the partition function of lattice QCD at infinite
coupling.  We could therefore apply the transformation to all links of
a lattice and then integrate out the fermions.  In this regard, the
most important property of the transformation \eqref{eq:uncolfla} is
that in terms of $\psi$ and $\bar\psi$, the right-hand side is local
in coordinate space.  Therefore, the fermion matrix, which for the
usual lattice gauge action is an irreducible matrix of dimension $\sim
N_cV$, decomposes into blocks of size $\sim N_f$ (number of flavors).
This means that the fermion determinant of the transformed action is
the product of de\-ter\-mi\-nants of fairly small matrices, which
bears promise for a new approach to fermion algorithms.

In order to apply this strategy to lattice QCD, we need to derive a
color-flavor transformation for gauge group SU($N_c$). This is done in
Sec.~\ref{sec:suN}. In Sec.~\ref{sec:applic}, we address the issue of
how to go beyond infinite coupling (i.e.\ how to include the plaquette
action), discuss baryon loops that arise in SU($N_c$), and comment on
simulation algorithms.

\section{Color-flavor transformation for SU($N_c$)}
\label{sec:suN}

We consider two types of tensors $\psi$ and $\varphi$ with only fermionic
degrees of freedom.  This simplifies the calculations and is more
appropriate for the application to lattice QCD that we have in mind.
The reader may think of $\bar\psi$ as corresponding to
$\bar\psi_{x+\hat\mu}$, $\psi$ to $\psi_x$, $\bar\varphi$ to
$\bar\psi_x$, and $\varphi$ to $\psi_{x+\hat\mu}$.  The result for the
transformation for SU($N_c$) is \vspace*{-2mm}
\begin{multline}
  \int_{\text{SU}(N_c)} dU
  \exp\left(\bar{\psi}_{{ a}}^{i}{ U^{ij}}
    \psi_{{ a}}^{ j}+\bar{\varphi}_{{ a}}^{ i}
    { U^{\dagger ij}}\varphi_{{ a}}^{ j}\right) \\ 
=C\cdot C_0\int_{\mathbb{C}^{N_f\times N_f}} 
\frac{d ZdZ^{\dag}}{\det(\1+{ ZZ^{\dag}})^{2N_f+N_c}}\\[-1mm]
  \times\exp\left(\bar{\psi}_{{ a}}^{ i} { Z_{ab}}
    \varphi_{{ b}}^{ i}-\bar{\varphi}_{{ a}}^{ i}
    { Z_{ab}^{\dag}}\psi_{{ b}}^{ i} \right)
  {\sum_{Q=0}^{N_f}} \chi_Q\: ,
\label{eq:suncolflav}
\end{multline}
where $a,b=1,\ldots,N_f$, $\chi_0=1$ and
\begin{equation}
\label{eq:chiQ}
\chi_{Q>0}=\mathcal{C}_Q
  \left[ \det(\mathcal{M})^Q+\det(\mathcal{ N})^Q\right].
\end{equation}
The matrices in the determinants in Eq.~\eqref{eq:chiQ} carry color indices.
They are 
$\mathcal{M}^{ij}=\bar{\psi}_{a}^{i}(\1+ZZ^{\dag})_{ab}\psi_{b}^{j}$
and
$\mathcal{N}^{ij}=\bar{\varphi}_{a}^{i}(\1+Z^{\dag}Z)_{ab}\varphi_{b}^{j}$.
Finally, the constants are given by
\begin{align}
  C&={\textstyle (1/\pi^{N_f^2})\prod_{n=0}^{N_f-1}(N_f+n)!/n!}
  \nonumber\\
  C_0&={\textstyle\prod_{n=0}^{N_f-1}n!(N_c\!+\!N_f\!+\!n)!/
    (N_c\!+\!n)!(N_f\!+\!n)!} \nonumber\\
  \mathcal{C}_Q &= \frac{1}{(Q!)^{N_c}(N_c!)^{Q}}
  \prod_{n=0}^{Q-1} 
  \frac{(N_c\!+\!n)!(N_f\!+\!n)!}{n!(N_c\!+\!N_f\!+\!n)!}\: .
\end{align}

The basic idea behind the proof of the color-flavor transformation is
the following. One can define an auxiliary Fock space, with fermion
creation and annihilation operators acting in this space. The
integrand of the SU($N_c$) integral can be regarded as a
complex-valued overlap of two states in this Fock space. Integration
over SU($N_c$) can be identified with a projection onto the
color-neutral sector of the Fock space. The idea is to develop another
implementation of this very projector, which then leads to the
color-flavor transformed integral. The details of the proof are rather
complicated, and we shall only sketch some of the steps involved. For
the full proof, we refer to Ref.~\cite{Schl02} (see also
Ref.~\cite{BNSZ}).

As mentioned above, we define two sets of fermionic creation and
annihilation operators $\bar{c}_{a}^{i}, c_{a}^{i}$ and $\bar{d}_a^i,
d_a^i$ which carry color indices $i=1,\ldots,N_c$ and flavor indices
$a=1,\ldots,N_f$. These operators obey canonical anti-commutation
relations.  The Fock space is obtained by acting on the Fock vacuum
$\ket{0}$ with linear combinations of the creation operators.

It is useful to define a certain ``half-filled'' state in this Fock
space as 
\begin{equation}
\ket{\psi_0}=(\bar{c}_{1}^{1}
    \cdots\bar{c}_{N_f}^{1})\cdots (\bar{c}_{1}^{N_c} \cdots
    \bar{c}_{N_f}^{N_c})\ket{0}\: .
\end{equation} 

The outline of the proof is as follows,
\begin{align} 
  &\int_{\text{SU}(N_c)} \!\!\!d{ U}
  \exp\left(\bar{\psi}_{{ a}}^{ i}{ U^{ij}}
    \psi_{{ a}}^{ j}+\bar{\varphi}_{{ a}}^{ i}
    { U^{\dagger ij}}\varphi_{{ a}}^{ j}\right) \nonumber\\
  &=\int_{\text{SU}(N_c)}\!\!\!d{ U}\bra{\psi_0}
  \exp\left(\bar{\psi}_{{ a}}^{ i} d_{{ a}}^{ i}-\bar{c}_{{ a}}^{ i} 
   \varphi_{{ a}}^{ i}\right) \nonumber\\[-2mm]
 &\hspace{2cm}\times\exp\left(\bar d_{{ a}}^{ i}{ U^{ij}}
    \psi_{{ a}}^{ j}+\bar{\varphi}_{{ a}}^{ i}
    { U^{\dagger ij}} c_{{ a}}^{j}\right)\ket{\psi_0} \nonumber
\end{align} 
\begin{align} 
  &=\bra{\psi_0} \exp\left(\bar{\psi}_{a}^{i} d_{a}^{i}
     -\bar{c}_{a}^{i} \varphi_{a}^{i}\right) \nonumber\\
  &\hspace{2cm}\times P\exp\left(\bar d_{a}^{i}
    \psi_{a}^{ i}+\bar{\varphi} _{a}^{i}
    c_{a}^{i}\right)\ket{\psi_0} \nonumber\\
  &={\sum_{Q=-N_f}^{N_f}} \bra{\psi_0}\exp\left( 
    \bar{\psi}_{a}^{i} d_{a}^{i}-\bar{c}_{a}^{i} \varphi_{a}^{i}\right) 
    \nonumber\\[-3mm]
  &\hspace{2cm}\times\1_Q\exp\left(\bar{d}_{a}^{i}\psi_{a}^{i}
    +\bar{\varphi}_{a}^{i}c_{a}^{i}\right)\ket{\psi_0} \nonumber\\[2mm]
  &= C_0\int \frac{D(Z,Z^{\dag})}{\det(1+ZZ^\dagger)^{N_c}} \nonumber\\[-2mm]
  &\hspace{0.6cm}\times\exp\left(\bar{\psi}_{a}^{ i}Z_{ab}
    \varphi_{b}^{ i}-\bar{\varphi}_{a}^{ i}
    Z_{ab}^{\dag}\psi_{b}^{i} \right)
  {\sum_{Q=-N_f}^{N_f}} \chi_Q\: .
\end{align}

The interpretation of integration over SU($N_c$) as a projection onto 
the color-neutral sector of Fock space can be understood by analogy with the
case of projecting a vector in $\mathbb{R}^3$ onto the $z$-axis.
By averaging over all SO($2$) rotations around the $z$-axis, what
remains is the $z$-component of the initial vector, which is precisely
the component that is invariant under such rotations.

In order to derive an alternative expression for the projector $P$, we
consider bilinear products of the creation and annihilation operators
of the form $\bar{c}_A^i c_B^j$. Here, $A,B=1,\ldots,2N_f$ are
composite indices introduced so that we don't have to distinguish
between the $c_a^i$'s and $d_a^i$'s. These bilinears generate a
gl($2N_fN_c$) algebra, which has two commuting subalgebras
that are of interest to us: the sl($N_c$) color algebra and the
gl($2N_f$) flavor algebra. The former is generated by operators
$\mathcal{E}^{ij}=\sum_A\bar{c}^i_A c^j_A$, $i\neq j$, and
$\mathcal{H}^i =\sum_A (\bar{c}^i_A c^i_A-\bar{c}^{N_c}_A c^{N_c}_A)$,
$i=1,\ldots, N_c -1$, and the latter is generated by $E_{AB}=\sum_i
(\bar{c}^i_A c^i_B-\frac12\delta_{AB})$.

States in the color-neutral sector satisfy
$T\ket{\mathcal{N}}=0$ for $T\in\text{sl}(N_c)$, or 
more explicitly,
\begin{equation}
{\textstyle\sum_{A=1}^{2N_f}\bar{c}_A^i c_A^j\ket{\mathcal{N}}
=(N_f+Q)\:\delta^{ij}\ket{\mathcal{N}}\:.}
\end{equation}
The possible values for $Q$ are $Q=-N_f,\ldots,N_f$.
Thus, the color-neutral sector splits into
subsectors labelled by $Q$. Under the action of U($2N_f$) on 
Fock space, given by
\begin{equation}
g\in\text{U}(2N_f)\; \mapsto \; 
T_g=\exp(\bar c_{A}^{i}(\log g)_{AB}c_{B}^{i})\:,
\end{equation}
each sector $Q$ subtends an irreducible representation of the flavor
group U($2N_f$) with a rectangular Young diagram that has $N_f+Q$ rows
and $N_c$ columns.

Using the method of generalized coherent states, we can express the
projector onto the color-neutral sector as a sum of projectors onto
the different subspaces $Q$, i.e.\ $P = \sum_{Q=-N_f}^{N_f} \1_Q$.

To derive the projectors $\1_Q$, we start out with the highest weight
vector of the representation $Q$, which is given by
$\ket{\psi_Q}= \prod_{i=1}^{N_c}\prod_{A=1}^{N_f+Q}
    \bar c_{A}^{i}\ket{0}$.
An overcomplete set of states is obtained by acting on this state with
all group elements $T_g,\; g\in \text{U}(2N_f)$. In the sector $Q=0$,
the highest-weight vector $\ket{\psi_0}$ corresponds to
``half-filling.'' Clearly, $T_h\ket{\psi_0}\propto\ket{\psi_0}$ for
$h=\diag(h_+,h_-)$ with $h_\pm\in\text{U}(N_f)$. The subgroup
$H=\text{U}(N_f)\times\text{U}(N_f)$ of U($2N_f$) is called the isotropy
subgroup of $\ket{\psi_0}$. In the sector $Q=0$,
coherent states may thus be parameterized without overcounting by the
elements of the coset space U$(2N_f)/H$. 
We label the elements of this coset space by picking a representative of
each equivalence class $gH$, $s(\pi(g))$. 
Each element $g\in\text{U}(2N_f)$ can be decomposed into a product 
$g=s(\pi(g))h(g)$.
As an explicit expression we choose $s(\pi(g))=$
\begin{equation}
\mat{\!(1+Z^{\dag}Z)^{-1/2}\!}{\!-Z^{\dag}(1+ZZ^{\dag})^{-1/2}\!}
    {\!Z(1+Z^{\dag}Z)^{-1/2}\!}{\!(1+ZZ^{\dag})^{-1/2}\!}\: ,
\end{equation}
where $Z$ is an arbitrary $N_f\times N_f$ complex matrix.
The action of $T_s$ on $\ket{\psi_0}$ can be shown to give
\begin{multline}
  \ket{Z}=\exp(\bar{d}_{a}^{i} 
  Z_{ab}  c_{b}^{i})
  \ket{\psi_0}\det(1+{Z^{\dag}}{Z})^{-N_c/2}, \\
\text{where } a,b=1,\ldots,N_f.\hfill
\end{multline}

The resolution of the identity in coherent states in the $Q$-sector is
given by
\begin{equation}
{\textstyle \1_Q = C_Q
  \int_{\text{U}(2N_f)}d\mu(g)\;T_g\ket{\psi_Q}\bra{\psi_Q}T_g^{-1}\:.} 
\end{equation}
In fact, this is a projector onto the $Q$-sector since it is easily seen
to annihilate all other states.

An integral over the whole group U($2N_f$) can be decomposed into
two integrals, one over $H=$U$(N_f)\times$U$(N_f)$, and one over
the coset space U$(2N_f)/H$,
where the measure of integration over the coset space is
given by $D(Z,Z^{\dag})=CdZdZ^{\dagger}/\det(1+{ Z}{ Z^{\dag}})^{2N_f}$.
For $Q=0$, the integrand is indeed independent of $h$, and with
the normalization vol$(H)=1$, the projector in this sector is simply
\begin{equation}
\1_0 = C_0\int D(Z,Z^{\dag})\ket{Z}\bra{Z} \:.
\end{equation}

The case of $Q\neq 0$ is more complicated, since the integration over
$H$ is non-trivial. However, this integration can be done using
standard group theory results, which finally leads to the
color-flavor transformed integral of Eq.~\eqref{eq:suncolflav}.

\section{Application to Lattice QCD}
\label{sec:applic}

\subsection{Induced QCD: How to generate the plaquette action}

The color-flavor transformation can be applied to each lattice link
for lattice actions at infinite coupling, i.e.\ without a plaquette
interaction term. Any program aimed at reformulating full QCD using
the color-flavor transformation must therefore address the issue of
including an interaction term for the gauge fields. Here, we take
advantage of the idea of induced QCD, which goes back to Kazakov and
Migdal \cite{Kaza92} and was further developed in Ref.~\cite{HasAP92}
leading to the result of interest to us. The idea is to couple a
number of additional heavy fermions to the gauge field.  This induces
gauge interactions, which can be directly related to an effective
(non-zero) gauge coupling.

To see how this works, we introduce $N_h$ heavy fermions, described by
the Wilson-Dirac operator \vspace*{-3mm}
\begin{equation}
D_{yx}=\delta_{yx}-\kappa\sum_{\mu=\pm1}^{\pm4}
\delta_{y,x+\hat\mu}(r+\gamma_\mu)U_\mu(x) \:,
\end{equation}
where we have not yet set $r=1$.  The hopping parameter
$\kappa=1/(2Ma+8r)$ tends to zero in the limit $M\to\infty$.  We now
integrate out the heavy quark fields to obtain the determinant of the
Wilson-Dirac operator, $D=\1-\kappa A$. The resulting expression can
then be expanded in powers of the hopping parameter,
\begin{align}
{\det}^{N_h} D &=\exp(N_h\tr\log(\1-\kappa A)) \nonumber\\
 &=\exp\left[-N_h\tr\left(\kappa A+{\textstyle\frac{\kappa^2}2} A^2
       \right.\right.\nonumber\\
 &\left.\left.\hspace{1.6cm}+{\textstyle\frac{\kappa^3}3} A^3
      +{\textstyle\frac{\kappa^4}4} A^4+\ldots\right)\right].
\end{align}
It is easily seen that terms containing odd powers of $A$ vanish when
one takes the trace, since the matrix then has no entries which are
diagonal in lattice site indices. The quadratic term is just a
constant (equal to zero for $r=1$), since
$U_{\mu}(x)U_{\mu}^{\dag}(x)=\1$.  Similarly, $\tr A^4=\tr
A_{xy}A_{yz}A_{zw}A_{wx}$ contains many constant terms as well as
terms of the form
\begin{align}
  \tr & \:\bigl[(r-\gamma_\nu)(r-\gamma_\mu)(r+\gamma_\nu)
  (r+\gamma_\mu)\nonumber\\
  & \hspace{1cm} U_\nu^\dagger(x)U_\mu^\dagger(x+\hat\nu)
  U_\nu(x+\hat\mu)U_\mu(x) \bigr]\nonumber\\
  =&-4(1+2r^2-r^4)\,\tr\, U_p \:.
\end{align}
Here, $p=(x;\mu\nu)$ denotes the lattice plaquette.  Note the overall
minus sign in the last line, which is essential to obtain the correct
continuum limit.  (In fact, with spinless fermions one obtains the
wrong sign.  The correct sign came from the lattice Lo\-rentz
structure of Wilson fermions, for any $0\le r\le1$.  Staggered
fermions also yield the correct sign, which arises from the staggered
phases.  However, with staggered fermions, it turns out that one needs
twice as many flavors to get the same coupling $g$, which
correspondingly enlarges the dimension of the matrix $Z$ in the
color-flavor transformed action.)

Collecting all relevant terms, we obtain at fourth order in $\kappa$
\begin{equation}
-8N_h\kappa^4(1+2r^2-r^4)\sum\limits_p\re\,\tr\, U_p\:,
\end{equation}
which is just the familiar plaquette action. We can thus identify
\begin{equation}
g^{-2}=4N_h\kappa^4(1+2r^2-r^4)\:,
\end{equation}
which becomes $8N_h\kappa^4$ for $r=1$.

In order to eliminate the effect of higher-order terms in the exponent
at fixed lattice spacing $a$, one can let $\kappa\to0$ and
$N_h\to\infty$ in such a way that $N_h\kappa^4=\text{const}$.  How
large $N_h$ has to be in practice to obtain the correct Yang-Mills
theory must be determined numerically.  A one loop calculation yields
$N_h>11N_c/2$ \cite{HasAP92}.  Incidentally, this implies that an
approach based on the color-flavor transformation together with
induced QCD is unlikely to be a suitable method for studying the
large-$N_c$ limit of Yang-Mills theory.

\subsection{Baryon loop expansion}

The color-flavor transformation deals with integrals that correspond 
to a single link of the lattice. In order to apply it to the lattice QCD
partition function, we need to apply it to all links. Furthermore, the
Dirac indices on $\psi$ and $\bar{\psi}$ have to be treated on par with
the flavor indices as far as the transformation is concerned. Thus, the
dimension of the $Z$ matrices obtained after the transformation is
given by $\dim(Z)=4(N_f+N_h)\equiv 4N_q$.

For Wilson fermions, the action at infinite coupling is given by
\begin{align}
S = &-\frac12\sum_{x,\mu>0} \left[
\bar{\psi}_{x+\hat{\mu},a}^{\alpha,i}\left( r+\gamma_{\mu}
\right)^{\alpha\beta}\!U_{x,\mu}^{ij}\:
\psi_{x,a}^{\beta,j}
\right. \nonumber
\\[-2mm] 
&\hspace{1.8cm}
+\left.
\bar{\psi}_{x,a}^{\alpha,i} \left( 
r-\gamma_{\mu}\right)^{\alpha\beta}\!
U^{\dagger
  ij}_{x,\mu}\:\psi_{x+\hat{\mu},a}^{\beta,j}
\right]  \nonumber\\
&+\sum_{x,a}(m_a+4r)\bar{\psi}_{x,a}^{\alpha,i}\psi_{x,a}^{\alpha,i}\:.
\end{align}
We have included all the indices, namely Greek Dirac indices
$\alpha,\beta = 1,\ldots,4$, color indices $i,j=1,\ldots,N_c$ and
flavor indices $a=1,\ldots,N_q$.  Note that the $\gamma_{\mu}$ are
traceless, idempotent matrices, which can be diagonalized by unitary
matrices,
\begin{equation}
\gamma_{\mu} = \Gamma_{\mu}\text{diag}(1,1,-1,-1)\Gamma_{\mu}^{\dag}\:.
\end{equation}
If we now set $r=1$, we have 
\begin{align}
\frac{1}{2}(r+\gamma_{\mu}) &= \Gamma_{\mu}P_{+}\Gamma_{\mu}^{\dag}
=\Gamma_{\mu}P_{+}P_{+}\Gamma_{\mu}^{\dag} \\
\frac{1}{2}(r-\gamma_{\mu}) &= \Gamma_{\mu}P_{-}\Gamma_{\mu}^{\dag}
=\Gamma_{\mu}P_{-}P_{-}\Gamma_{\mu}^{\dag} \:.
\end{align}
Here, $P_{+}\!=\!\text{diag}(1,1,0,0)$ and $P_{-}\!=\!\text{diag}(0,0,1,1)$
are projection operators in Dirac space. On each link, we define
\begin{align}
\bar{\phi}_{x,a}^{\gamma,i} &\equiv \bar{\psi}_{x,a}^{\alpha,i}
\Gamma_{\mu}^{\alpha\beta}P_{-}^{\beta\gamma}\:,& 
\bar{\phi}_{x+\hat{\mu},a}^{\gamma,i}&\equiv
\bar{\psi}_{x+\hat{\mu},a}^{\alpha,i}\Gamma_{\mu}^{\alpha\beta}
P_{+}^{\beta\gamma}\:,\nonumber\\
\phi_{x,a}^{\gamma,i}&\equiv P_{+}^{\gamma\delta}
\Gamma_{\mu}^{\dag\delta\epsilon}\psi_{x,a}^{\epsilon,i}\:,&
\phi_{x+\hat{\mu},a}^{\gamma,i}&\equiv P_{-}^{\gamma\delta}
\Gamma_{\mu}^{\dagger\delta\epsilon}\psi_{x+\hat{\mu},a}^{\epsilon,i}
\:.\nonumber 
\end{align}
This definition is of course not the only possible choice for the
inclusion of the Dirac matrices, and therefore the color-flavor
transformed action will not be unique.  This should, however, have no
influence on the physics, since the transformation is in any case
still exact.  The $\bar{\phi}$ and $\phi$ above are still Grassmann
variables, and so we can now apply the color-flavor transformation to
each link of the lattice to obtain a transformed partition function of
the form
\begin{align}
Z_\text{QCD}=\prod_{\underset{\mu>0}x}&\int\frac{d{ Z}_\mu(x)
  d{Z}_\mu^{\dagger}(x)}{\det(1+{Z}_\mu(x){Z}_\mu^{\dagger}(x))^{8N_q+N_c}}
  \nonumber\\
  \times&\int D\bar\psi D\psi\;e^{-S_\mu(x)}
  \sum_{Q=0}^{4N_q}\chi_{Q\mu}(x)\: .
\label{eq:zcolfla}
\end{align}
In the exponential, there is now a local term 
$ S_\mu(x)=\bar\psi_{x,a}^{\alpha,i}B_\mu(x)_{ab}^{\alpha\rho} 
\psi_{x,b}^{\rho,i}$, where
\begin{multline}
 B_\mu(x)_{ab}^{\alpha\rho}=
[\Gamma_{\mu}^{\alpha\beta}P_{-}^{\beta\gamma}
Z_{x,\mu}^{\dag\gamma\delta;ab}
P_{+}^{\delta\epsilon}\Gamma_{\mu}^{\dag \epsilon\rho}] \\
    -[\Gamma_{\mu}^{\alpha\beta}P_{+}^{\beta\gamma}
 Z_{x-\hat{\mu},\mu}^{\gamma\delta;ab}P_{-}^{\delta\epsilon}
\Gamma_{\mu}^{\dag\epsilon\rho}]+(m_a+4)\delta_{ab}\delta^{\alpha\rho}\:.
\label{eq:fermdet}
\end{multline}
Furthermore, we have $\chi_{0\mu}(x)=1$ and
\begin{equation}
\chi_{Q\mu}(x)=\mathcal{C}_Q\left[{\det}^Q\mathcal{ M}_\mu(x)
    +{\det}^Q\mathcal{{ N}}_\mu(x)\right]\: .
\end{equation}
The matrices which appear here are defined as
\begin{align}
&\mathcal{M}_\mu^{ij}(x)=\nonumber\\
&\;\;\bar{\psi}_{x+\hat{\mu},a}^{\alpha,i}
\Gamma_{\mu}^{\alpha\beta}P_{+}^{\beta\gamma}
(1+Z_{x,\mu}Z_{x,\mu}^{\dag})^{\gamma\delta;ab}
P_{+}^{\delta\epsilon}\Gamma_{\mu}^{\dag\epsilon\rho}\psi_{x,b}^{\rho,j}
\nonumber\\
&\mathcal{N}_\mu^{ij}(x)=\nonumber\\
&\;\;\bar{\psi}_{x,a}^{\alpha,i}
\Gamma_{\mu}^{\alpha\beta}P_{-}^{\beta\gamma}
(1+Z_{x,\mu}^{\dag}Z_{x,\mu})^{\gamma\delta;ab}
P_{-}^{\delta\epsilon}\Gamma_{\mu}^{\dag\epsilon\rho}
\psi_{x+\hat{\mu},b}^{\rho,j} 
\: . \nonumber
\end{align}

The next step is to integrate out the fermions. This is 
straightforward for $Q=0$. With the definition of $B_{\mu}(x)$
from above, we define $B(x)=\sum\limits_\mu B_\mu(x)$ and
$B=\!\!\!\!\!\underset{\text{all sites $x$}}{\bigotimes}\!\!\!\!\!B(x)$. Then
$B$ has a block-diagonal structure, with different blocks
corresponding to different lattice sites. After integration
over the Grassmann variables, we thus obtain 
\begin{equation} 
Z^{Q=0}_{\rm QCD}= 
{\det}^{N_c}B=\prod\limits_\text{all sites $x$}  
{\det}^{N_c}B(x)\: . 
\end{equation} 
For $Q=0$, the situation becomes more complicated, since the
$\chi_{Q>0}$ terms induce non-local contributions to the action.
Notice that these terms correspond to baryon propagation, which
explains why they were absent in the case of gauge group U($N_c$).

The case of one static baryon ($Q=1$) was studied in saddle-point
approximation (for large $N_c$) in 1+1 dimensions by Budczies et al.\ 
in Ref.~\cite{BNSZ}.  However, our goal is an exact (numerical)
approach to QCD, which takes into account all possible values of $Q$.

In order to evaluate terms from $\chi_{Q>0}$ contributions,
we need to apply Wick's theorem. This results in terms involving
the inverse of the matrix $B$. Note that $B$ is diagonal in space 
and color, so that we have
\begin{equation}
(B^{-1})_{xy,{pq}}^{ij}=\delta^{ij}\:\delta_{xy}\:B^{-1}(x)_{pq}\:,
\end{equation}
where $p,q=1,\ldots,4N_q$ combine flavor and Dirac indices.  Let us
denote
\begin{equation}
    \sum_{Q=1}^{4N_q}\chi_{Q\mu}(x) \equiv 
    \stanlink{>}+\stanlink{<}
\end{equation}
Here, a solid link denotes the presence of $1,2,\ldots,4N_q$ baryons.
Now consider
\begin{equation}
\prod_{x,\mu}\Bigl(1+\stanlink{>}+\stanlink{<}\;\Bigr)
\end{equation}
We can expand the product and note that because of
\blm{(B^{-1})_{xy}\sim\delta_{xy}}, only those terms survive that
  correspond to closed loops (cf.\ for example Fig.~\ref{fig:plaqexpand}).
\begin{figure}[tb]
\begin{center}
\setlength{\unitlength}{0.15mm}
\vspace*{-3mm}
\begin{multline*}
\left(1+\plaq{b}\right)\left(1+\plaq{r}\right) 
\left(1+\plaq{t}\right)\left(1+\plaq{l}\right)\\
=1+\plaq{a}\phantom{WWWWWWW}
\end{multline*}
\end{center}
\vspace*{-10mm}
\caption{Expansion of baryon contributions: Only closed loops survive
  the integration over the Grassmann variables. \vspace*{-5mm}
\label{fig:plaqexpand}}
\end{figure}
Hence, in the partition function only closed baryon loops are allowed.
These must not backtrack, so that each link occurs at most once in a
graph. The paths may self intersect, and disconnected loops are also
allowed.  In addition, for each site the total number of baryons on
the links connected to this site must not exceed $4N_q$.

Let us now consider observables, such as correlation functions.
(In the color-flavor transformed formulation, gauge fields no longer
appear, and it is therefore not surprising that non-gauge-invariant 
quantities are manifestly equal to zero.) As an example, take
the baryon two-point function 
$\langle\bar{\mathcal{B}}_y{\mathcal{B}}_z\rangle$, 
with $\mathcal{B}=\varepsilon_{ ijk}\psi^{i}\psi^{ j}\psi^{ k}$.
Schematically, we have to compute
\begin{equation}
\bar{\mathcal{B}}_y \mathcal{B}_z\prod_{x,\mu}
    \Bigl(1+\stanlink{>}+\stanlink{<}\;\Bigr)
\end{equation}
Contributing baryon paths are those which originate at the site $y$
and terminate at site $z$.
Of course, a background of closed baryon loops, such as those appearing
for the partition function, is also allowed, in addition to the baryon
propagating from $y$ to $z$.

\subsection{Algorithmic considerations}

While it is, in principle, clear how to perform a simulation for the
$Q=0$ case, the presence of baryon loops poses additional problems.
Clearly, the number of possible loops grows prohibitively with the
lattice volume so that a probabilistic approach to the generation of
such loops is needed. A possibility is some form of random walk
algorithm \cite{SS}. It should be noted, though, that the coefficients
$\mathcal{C}_Q$ decrease rapidly as functions of $Q$ so that
multi-baryon paths are potentially strongly suppressed.  For example,
for $Q=1$ and $N_c=3$, we have $\mathcal{C}_1\propto1/N_q^3$, which is
already much smaller than in the $Q=0$ case.  In addition, since
longer loops involve more factors of $\mathcal{C}_Q$, only small
loops, or shortest paths in the case of correlation functions, will
significantly contribute. An expansion in baryonic loops may therefore
be a far more systematic and better controlled approach than
quenching, for example.

Unfortunately, the color-flavor transformed action is complex. 
This is already manifest for $Q=0$, as
can be seen from Eqs.~\eqref{eq:zcolfla} and \eqref{eq:fermdet}, since
$Z$ is a general complex matrix.
We have tried to attack this problem by partially integrating out some
of the degrees of freedom of $Z$ (cf.\ Ref.~\cite{ScWe02}), and also by
finding different parameterizations of the coset space
U($2N_f$)/[U($N_f$)$\times$U($N_f$)], but so far have not been able to
obtain a satisfactory result. However, some progress has been made
recently in attacking fermion sign and complex action problems
\cite{Chan99}, and this may also turn out to be helpful for the
color-flavor transformation.

\section{Summary and Outlook}

We have derived the color-flavor transformation for gauge group
SU($N_c$).  This transformation converts a certain type of integral
over color gauge fields to an integral over flavor matrices. The
integral being transformed is of the same form as the one encountered
on each link of the lattice QCD partition function at infinite
coupling. Since (for zero baryons) the fermion determinant after the
transformation is block-diagonal in the lattice site indices, this
approach has the promise of leading to new ways of simulating
dynamical fermions. However, in contrast to gauge group U($N_c$), the
SU($N_c$) case gives rise to additional baryon loops, which lead to
nearest-neighbor coupled fermion terms. Nevertheless, multi-baryon
loops are potentially suppressed, and a systematic expansion of the
partition function in terms of such loops is possible.

In order to go beyond the infinite coupling limit, a number $N_h$ of
heavy flavors can be added to the theory to generate the plaquette
action ($g^{-2}=8N_h\kappa^4$ for Wilson fermions).

In the present formulation, the color-flavor transformed action is
complex, which makes numerical simulations based on this approach
impractical without further improvements. Work in this direction is in
progress. 

\medskip

{\bf Acknowledgments}. This research was supported in part by DOE
contract No.\ DE-FG02-91ER40608 and by a Hellman Family Fellowship.

\end{document}